\documentclass[conference]{IEEEtran}

\usepackage{graphicx}
\usepackage{caption}
\usepackage{subcaption}

\begin{document}
\title{Quantum Security for the Physical Layer}

\author{
\IEEEauthorblockN{Travis S. Humble}
\IEEEauthorblockA{Quantum Computing Institute\\
Oak Ridge National Laboratory\\
Oak Ridge, Tennessee 37831-6015\\
Email: humblets@ornl.gov}
}

\maketitle

\begin{abstract}
The physical layer describes how communication signals are encoded and transmitted across a channel. Physical security often requires either restricting access to the channel or performing periodic manual inspections. In this tutorial, we describe how the field of quantum communication offers new techniques for securing the physical layer. We describe the use of quantum seals as a unique way to test the integrity and authenticity of a communication channel and to provide security for the physical layer. We present the theoretical and physical underpinnings of quantum seals including the quantum optical encoding used at the transmitter and the test for non-locality used at the receiver. We describe how the envisioned quantum physical sublayer senses tampering and how coordination with higher protocol layers allow quantum seals to influence secure routing or tailor data management methods. We conclude by discussing challenges in the development of quantum seals, the overlap with existing quantum key distribution cryptographic services, and the relevance of a quantum physical sublayer to the future of communication security.
\end{abstract}

\section{Introduction}
\label{sec:intro}
The physical layer is the lowest layer in any communication protocol suite. Often abbreviated as PHY, it represents specifications for generating, modulating, and transmitting signals over a network. Securing the PHY layer, however, is notably different from the other layers comprising the protocol stack. Higher protocol layers resort to security schemes that naturally lie within the cyber domain, such as encryption or digital signatures, while similar capabilities for the physical domain have not yet been developed. This is due to the fundamentally different type of security required for controlling access to a physical medium.
\par
The security of the physical layer is typically managed by restricting access to its constituent pieces. For example, fiber buried underground is usually assumed to be inaccessible -- although this method of management is not reliable since a buried fiber can be easily exposed by digging it up or accessed through its service points. Many networks have large spatial footprints and frequent visual inspection is simply impractical. Automated diagnostics such as spectral analysis or optical time-domain reflectrometry check fiber health, but these methods are useful when the network is inactive and are not intended for real-time, inline monitoring. In addition, these methods require verified reference traces to validate each link, and the number of those traces grows quickly with network connectivity. 
\par
Alongside these practical hurdles, the physical layer also exhibits a fundamental vulnerability that remains unanswered by present security practices. The modern view of {PHY} has been built upon the principles of classical physics, notably, classical electromagnetic theory for preparing and transmitting signals. This fundamental feature, although necessary, also provides an attacker the opportunity to compromise the integrity of a transmission. That is to say, by measuring the frequency, bandwidth, and modulation of a transmission, an attacker can resend a replica without revealing their presence. This vulnerability, often described as an intercept-resend attack or the man in the middle attack, derives directly from the implementation of a classical physical layer.
\begin{figure}[h]
\begin{center}
\includegraphics[scale=0.4]{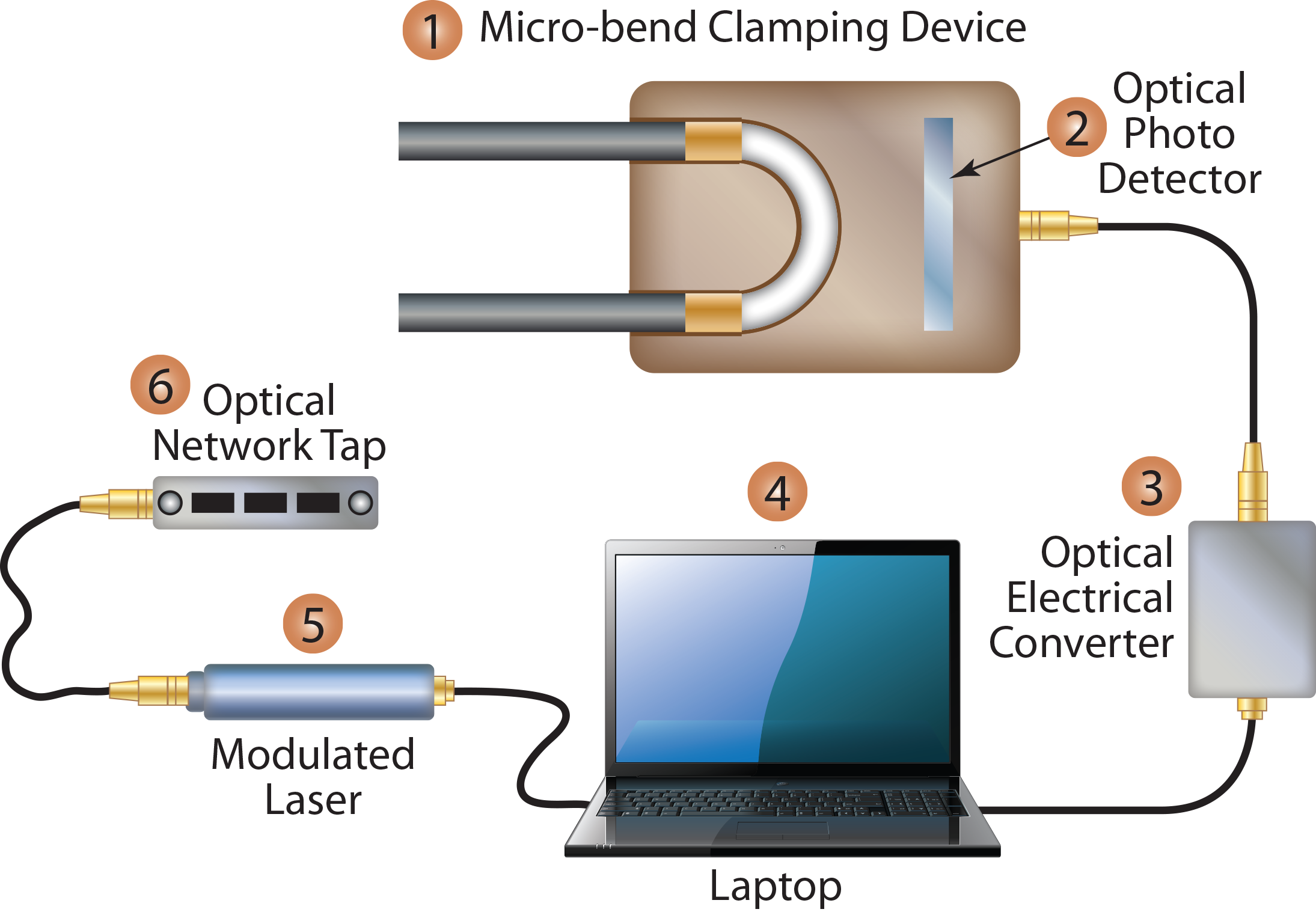}
\caption{The intercept-resend or man-in-the-middle attack can be implemented either by intercepting the signal using a network tap or by monitoring leakage from a bent fiber. The optical signal can then be sampled and analyzed. The knowledge acquired about the signal can then be used to craft a replica pulse that is injected into the fiber. The basis for this vulnerability is the exact cloning of signals in classical mechanical theory.}
\vspace{-10pt}
\label{fig:attack}
\end{center}
\end{figure}
\par
It was once reasonable to assume the technological sophistication required for intercepting and transmitting light prohibited the intercept-resend attack at the physical layer. However, advancements in both optical network technology and computing power have greatly increased the feasibility of this attack at relatively low cost to the attacker. A shown in Fig. \ref{fig:attack}, it is straightforward to intercept an optical signal using an optical-electrical converter and to resend the same or different signal using an optical transmitter. An attacker may also use existing optical technology such as a repeater, multiplexer, or fiber network tap to perfectly replicate the transmitted light or use optical transmitter technology alongside {\it a priori} knowledge of the pulse encoding to inject a spoofing pulse. Any of these methods can be used to exploit existing fiber in the physical layer.
\par
Quantum communication offers a means of securing the physical layer by closing the man in the middle attack. This capability can not be provided by conventional communication but instead stems from the indeterminism underlying quantum physics. The technology needed to take advantage of these capabilities is now being realized and brought to commercial markets. We expect that these components will be useful for providing security for the physical layer as part of future communication protocols. In this contribution, we show how quantum seals can be realized and applied to securing the physical layer.  The presence of a quantum physical sublayer within the network would provide a unique instance of a cyber-physical systems that could be used to develop secure communication services.
\par
The paper is organized as follows. Following the Introduction of Sec.~\ref{sec:intro}, we present the Principles of Quantum Communication  in Sec.~\ref{sec:principles}, which defines the basic nomenclature and physical principles. This followed by an introduction to Quantum Optical Communication in Sec.~\ref{sec:optics}, which covers the technological basis for a quantum seal. We next discuss the operation and behavior of quantum seals in Sec.~\ref{sec:seals} before describing their  integration into a cyber-physical systems in Sec.~\ref{sec:cps}. That section also includes a description of how the security of the physical layer can play a role in the behaviors of higher-protocol layers and secure communication services. Finally, we summarize the outlook for quantum-enabled security based on a quantum physical sublayer in Sec.~\ref{sec:out}.
\section{Principles of Quantum Communication}
\label{sec:principles}
Quantum communication is rooted in the field of quantum information, which also includes quantum computing, quantum cryptography, and quantum sensing. At the heart of these disciplines is the use of quantum mechanics as the underlying mathematical framework. This distinction has been shown to lead to several notable differences as compared to conventional information technologies based on classical mechanics \cite{Wilde2013}. In this section, we offer a primer on some of the principles of quantum communication and especially those that may be used for securing the PHY layer in a communication system.
\par
Quantum communication, like classical communication, is built up from logical states, namely, 1's and 0's. These logical states are the information that must be encoded into a physical system and relayed according to some chosen protocol. What distinguishes quantum information from classical information is how this encoding is accomplished. Conventional communication transceivers rely on classical signals to encode the logical states, e.g., positive and negative voltages, electromagnetic carrier waves, etc. These signals are well described by classical physics. Quantum communication differs as the physical systems used to encode bits of information behave according to the laws of quantum mechanics. This distinction has proven to have dramatic consequences.
\par
\begin{figure}
\centering
\includegraphics[width=.5\linewidth]{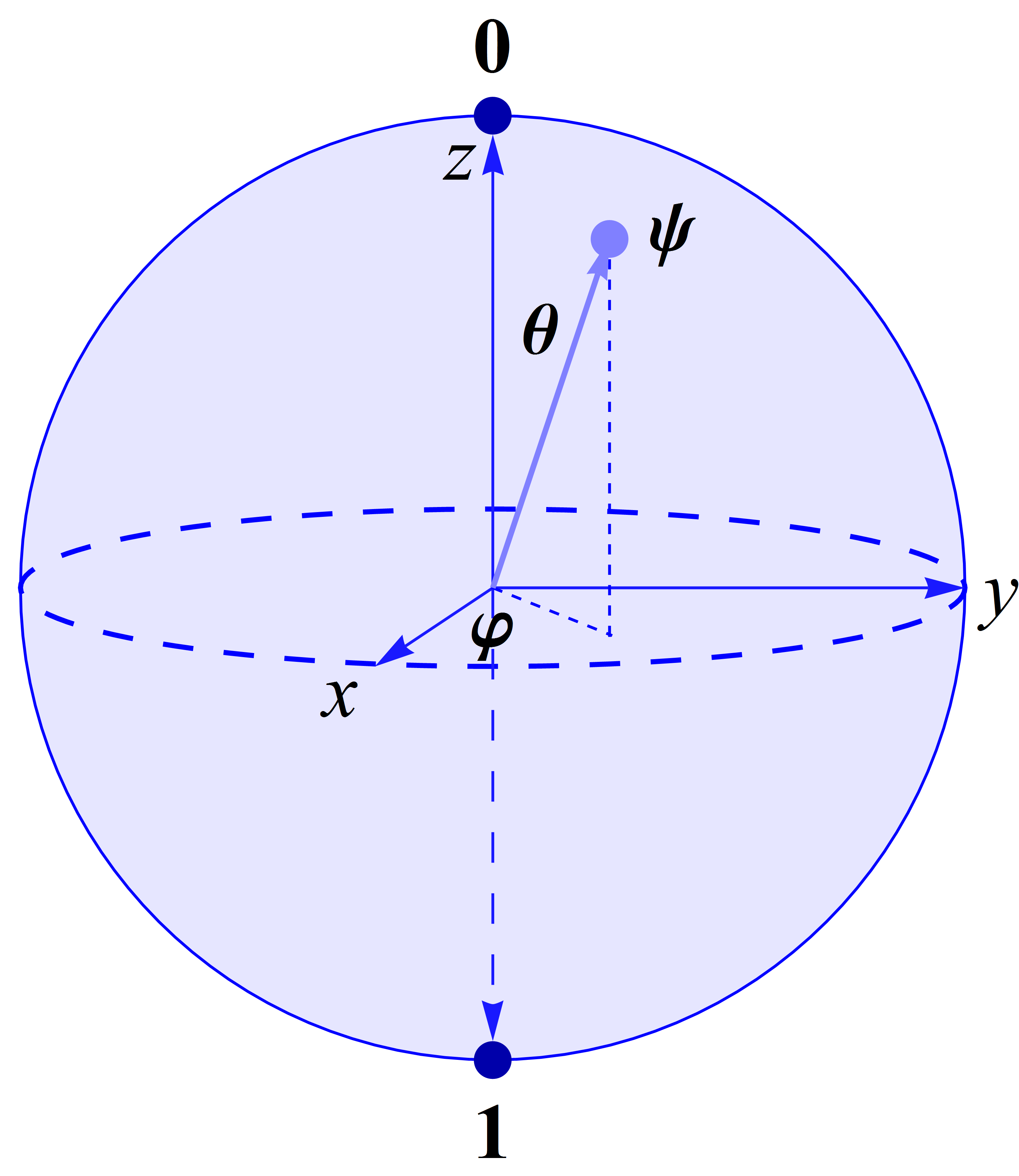}
\caption{The possible values of a single qubit map onto the unit sphere, also known as the Bloch sphere. The upper and lower poles of the sphere correspond to the logical states 0 and 1 respectively. A classical bit can only encode either the logical 0 or 1 state, while a qubit $\psi$ can encode any of the infinite number of superposition states specified by the angles $\theta$ and $\varphi$. }
\label{fig:comp}
\end{figure}
Unlike classical bits, quantum bits, i.e., \textit{qubits}, can exist in superpositions of logical states. This means that qubits need not encode a single value of either 1 or 0, but that they can simultaneously represent superpositions of both values.  This difference is highlighted in Fig. \ref{fig:comp}. The surface of the unit ball represents the set of all possible superpositions that a single qubit can encode. It is also known as the Bloch sphere within the quantum information community. On the Bloch sphere, the poles correspond with classical limits of 1 and 0 and represent the only possible states for a binary system, i.e., a bit. By contrast, the value $\psi$ of a qubit corresponds to one of the infinite points on the spherical surface. This value can be formalized by using the spherical poles as a two-dimensional basis and  the longitudinal and azimuthal angles $\theta$ and $\varphi$ to define the qubit value. There is an infinite number of these values and, consequently, an infinite number of possible qubit values. 
\par
The availability of an infinite set of possible qubits values does not imply that an infinite amount of information can be transmitted. It was first shown by Holevo that there is a limit on how much information can be communicated using a qubit and this maximum is precisely one bit per transmitted qubit \cite{Wilde2013}. This bit of information is retrieved from a qubit whenever a measurement is made. The measurement effectively projects the qubit into one of the two poles of the Bloch sphere. However,  which of these two outcomes is observed depends only partly on the value of the qubit being measured. For example, when the incoming qubit has amplitudes $c_0 = \cos\theta$ and $c_1 = \sin\theta$ with respect to the polar basis and $\varphi=0$, then the probabilities that the receiver records 0 and 1 are given respectively as $|c_0|^2$ and $|c_1|^2$. The magnitudes only describe the probability of a given observation - the actual outcome is indeterminate prior to the measurement. In the case of either outcome, however, the Holevo bound ensures the qubit $\psi$ provides only a single bit of information.
\par
In addition to individual qubits, quantum communication also permits preparing superpositions of multiple logical states. The most straightforward example is to consider a pair of qubits, each individually prepared. Qubits in this form are said to be  separable as each system can be described independently from the other. Separable qubits, however, do not exhaust the set of multi-qubit states permissible within quantum communication. There are quantum states for two qubits which can not be expressed in separable form. This property is the formal definition of \textit{entanglement} and it has been hailed as \textit{the} hallmark of quantum mechanics.  Forming superpositions of individual and multiple qubits is an important resource for quantum communication and will underlie our discussion of how quantum seals operate.
\par
One of the most notable consequences of entanglement is the apparent contradiction it poses to local realism, a central tenet of classical physics. The contradiction arises from the seemingly non-local correlations that can exists within entangled states. Local correlations between both classical and quantum bits are common, e.g., two well-defined transmissions may produce correlated outcomes. But entangled qubits also give rise to correlations which violate the common sense idea that disparate, random (probabilistic) events can not be strictly correlated. Even though individual measurement outcomes are unpredictable, as described above, the combined outcomes from a pair of entangled qubits can be perfectly correlated. That this is true even when the qubits are very remote from each other led to Albert Einstein's infamous description of ``spooky action at a distance''. It was only later that John Bell developed a framework to test the validity of local realism with quantum mechanics. Commonly termed Bell tests, these statements give upper limits for the strength of correlations that any local theory can exhibit. The classical limits set by these tests have been routinely violated by experiments using quantum mechanical systems including entangled photon pairs. We will invoke a similar test as the diagnostic measure of the quantum seal in Sec. ~\ref{sec:seals}.
\section{Quantum Optical Communication}
\label{sec:optics}
Realizing the novel aspects of quantum communication requires adopting an encoding in some selected quantum physical degree of freedom. As noted above, qubits store a continuously variable value that mimics an analog signal up to the time of measurement at which point it is converted into a digital domain. Modulation corresponds with preparation of the physical modes in a desired sequence just as is done with conventional communication. Differences arise in how measurements effect these modes and the correlations that can persist between disparate signals. In this section, we describe some of the ways that optical technology can be used to realize quantum communication.
\par
\begin{figure}[h]
\begin{center}
\includegraphics[scale=0.4]{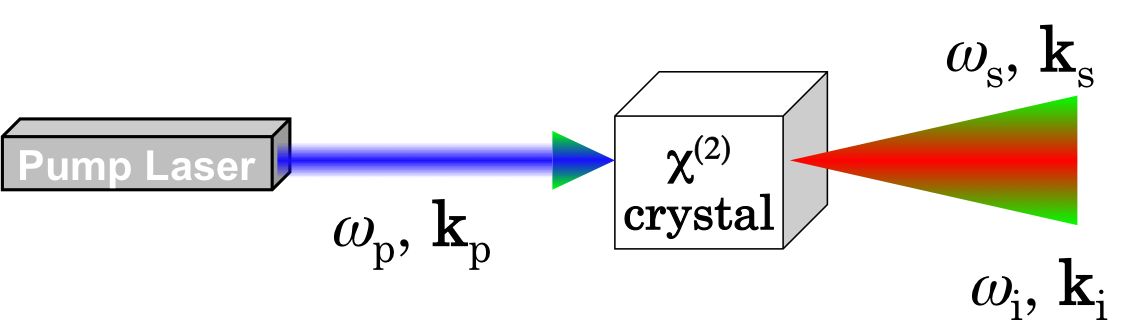}
\caption{An illustration of the spontaneous parametric down conversion (SPDC) process emphasizing the individual components in a bulk-crystal source. A pump laser with frequency $\omega_{p}$ and longitudinal wave vector $\mathbf{k}_p$ passes through a bulk crystal that has a non-zero second-order non-linearity $\chi^{(2)}$. The nonlinear crystal mediates the down conversion of a pump photon into a pair of lower-energy signal and idler photons. The signal and idler respectively have frequencies $\omega_s$ and $\omega_i$ and wave vectors $\mathbf{k}_{s}$ and $\mathbf{k}_{i}$, where the conservation of energy and momentum ensures $\omega_p = \omega_s + \omega_i$ and $\mathbf{k}_{p} = \mathbf{k}_{s} + \mathbf{k}_{i}$.}
\vspace{-10pt}
\label{fig:spdc}
\end{center}
\end{figure}

Over the past twenty years, individual photons have served as the most prominent medium for quantum communication. This use stems, in part, from the relative maturity of quantum optical technology as well as its suitability for communication applications. Techniques for generating single photons, both on demand and probabilistically, have been the focus of intense development  \cite{Eisaman2011}. As an example of these photon sources, the process of spontaneous parametric down-conversion (SPDC) has seen widespread use for quantum communication applications. In the SPDC process, a high-energy pump pulse composed from a macroscopic number of photons interacts with a nonlinear optical medium. The medium mediates an interaction in which the high-energy pump photon quickly decays into a pair of lower energy photons typically termed the signal and idler. A schematic of a basic experimental setup using a nonlinear optical crystal as the down conversion medium is shown in Fig. \ref{fig:spdc}. The overall rate of this process is exceptionally small with a probability on the order of $10^{-15}$. For a macroscopic number of pump photons, this process can be tuned to ensure with high probability that one photon is down converted from each pump pulse. The resulting signal and idler photons can then be used individually to encode single qubits of information or used as a pair to encode an entangled state.
\par
The photon, as a quantum of light, provides a variety of different physical properties that can be used to encode quantum information. This includes the spectral and spatial modes as well as the polarization, transverse momentum, orbital angular momentum, photon number, and field-quadrature variables. In general, the quantized levels of each photonic degree of freedom is appropriate for encoding. Polarization, for example, can encode a 0 using the horizontal state and a 1 with the vertical state. A polarization-encoded qubit is represented by the superposition of these different polarizations. 
\par
\begin{figure}[h]
\begin{center}
\includegraphics[scale=0.4]{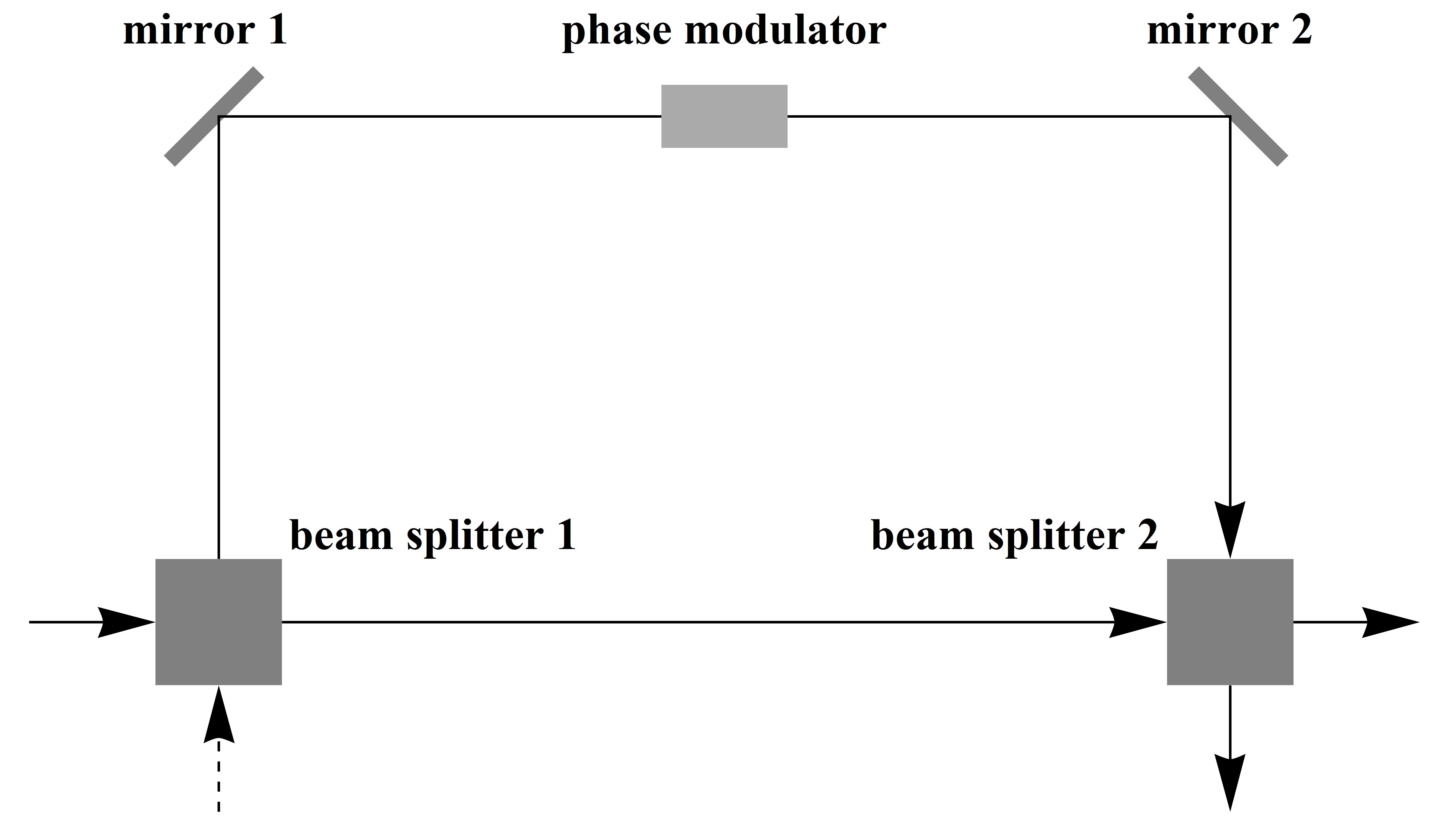}
\caption{An unbalanced Mach-Zehnder interferometer for preparing a super-position of arrival times for a single photon. The interferometer is constructed from a pair of 50:50 (non-polarizing) beam splitters and two mirrors, which are arranged to support one short path and one long path.  These two distinct spatial modes encode the binary basis for a qubit. While beam splitter 1 separates the photon into two paths, beam splitter 2 recombines them. The two pathways may interfere at the second beam splitter depending on the relative lengths and the value of the applied phase in the longer path.}
\vspace{-10pt}
\label{fig:MZI}
\end{center}
\end{figure}

Similarly, the spatial mode a photon occupies can encode a logical value. An example is shown using the unbalanced Mach-Zehnder interferometer in Fig.~\ref{fig:MZI}, where a single photon passes through a beam splitter. The lossless beam splitter operates by transmitting some portion of light while reflecting the remainder. These binary path labels can be used to encode 0 and 1. For a single photon, this amounts to a probability to either transmit or reflect through the beam splitter. When these probabilities are both 50\%, the photon is prepared in an equal superposition of the transmitted and reflected modes while the superposition state serves as a qubit. The two values can be distinguished based on the arrival time of the photons. In Fig.~\ref{fig:MZI}, a second beam splitter interferes the two photon prior to measurement. We will see in the next section that  interference plays a crucial role in testing for entanglement.
\par
It is also possible to construct quantum optical sources that provide good approximations to single-photon qubits by using strongly attenuated laser pulses. A laser emits a good approximation to a coherent beam of light that has a near Gaussian shape in the phase-space defined by the field quadratures. In the so-called `weak-pulse' regime, the coherent output from a conventional laser is heavily attenuated such that the average number of photons per mode is much less than one. This has the effect of reducing the strength of the field, which is proportional to photon number, and draws the state closer to the vacuum mode located at the origin. Because the average energy is below the single-photon level, measurement statistics for a weak pulse  provide a fair approximation to what is observed using individual photons. 
\section{Quantum Seals}
\label{sec:seals}
There are several notable uses for quantum communication. These include quantum key distribution (QKD) for establishing secure communications, quantum teleportation for busing information within a quantum computer, entanglement swapping for setting up long-distance quantum networks, and quantum seals for monitoring the integrity of physical boundaries. While this section focuses on the operation of quantum seals, many of the same principles of quantum communication and quantum optical encoding arise in the discussion of these other application domains.
\par
Seals are widely used for verifying the integrity of enclosed systems, including storage containers, physical perimeters, and fiber networks. In the optical context, fiber-optic seals are especially useful for actively surveying large areas or inventories \cite{Johnston2001, Szustakowski2005, Juarez2005}. These seals typically operate by confirming transmission of a classically encoded light pulse from source to receiver, where tampering is indicated by either the absence of the light or an error in the received encoding. 
\par
In the classical setting, detection of tampering assumes the intruder is unable to accurately replicate the original transmission. But if the attacker is able to perfectly replicate the classical light, e.g., using either an optical repeater or {\it a priori} knowledge, then classical instances of an optical seal are vulnerable to intercept-resend spoofing, i.e., the man-in-the middle attack. This vulnerability represents the attacker's ability to measure the classical information, such as the frequency, bandwidth, and modulation, that describes the state of the light and to resend a corresponding replica. These capabilities are well established in existing optical communication networks, which make extensive use of optical repeaters, amplifiers, and multiplexers for seamlessly replicating signals.
\par
The vulnerability of classical signals to a man-in-the-middle attack can be closed using a novel application of quantum information. The basis for this ``patch'' is the no-cloning theorem. The no-cloning theorem states that any attempt to clone information describing a quantum physical system necessarily introduces noise \cite{Wootters1982}. The consequence of the theorem is that attempts to intercept and resend a replica of a quantum system, e.g., an individual photon, are guaranteed to fail.  More important, the necessary imperfections of the man-in-the-middle attack appear prominently in later measurements made along the network. These quantum statistical anomalies can be quickly discovered and their significance quantified as part of a cyber-physical monitoring system.
\par
The guarantees of the no-cloning theorem also underlie the security offered by quantum key distribution (QKD). In QKD, the non-local correlations inherent to quantum states are used to establish correlated measurements outcomes between users. However, any attempt by an eavesdropper to clone the transmitted state introduces additional noise into the observed measurements that reveal her presence to the users. Identifying the eavesdropper requires a series of two-way communication to compute a quantum bit-error rate that serves as a diagnostic measurement of the physical layer. If categorized as being secure, i.e., if the error is low enough, then the users continue to negotiate the final secret key.
\begin{figure}[h]
\begin{center}
\includegraphics[scale=0.6]{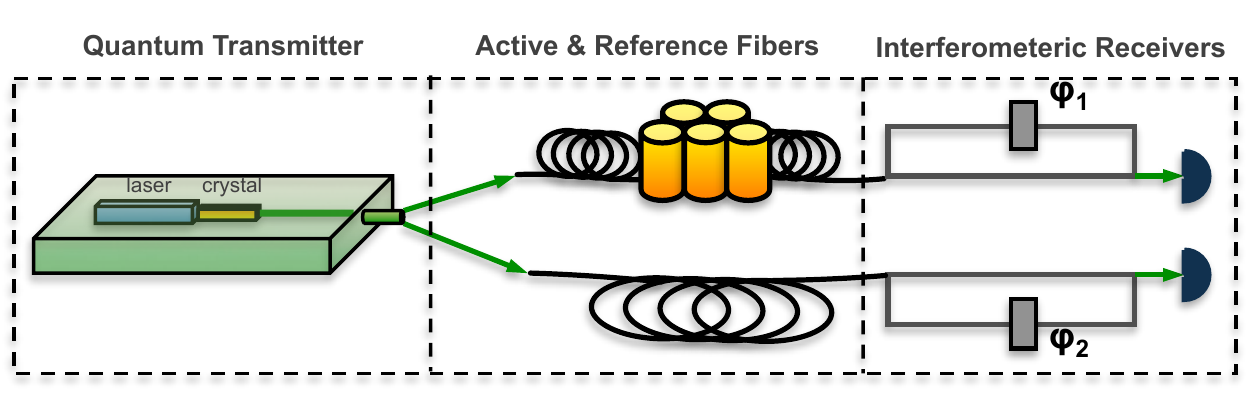}
\caption{An implementation of a quantum seal to monitor critical infrastructure. The seal is composed from a quantum transmitter based on SPDC that emits a pair of photons into the active and reference fibers. Each fiber terminates at an unbalanced Mach-Zehnder interferometer. The interferometric receivers measure the photon arrival times and a processing system monitors the correlations. Attempts by an intruder to replicate signals in the active fiber destroy the expected non-local correlations. The loss of entanglement manifests as a reduction in the measured correlations. When the measured correlations fail to exceed the threshold set by the Bell test, it is an indication that the seal has been compromised.}
\vspace{-10pt}
\label{fig:seal}
\end{center}
\end{figure}
\par
Quantum seals offers a more straightforward application of the no-cloning theorem. A diagram of the seal operation is shown in Fig. \ref{fig:seal}. The implementation describes a fiber-based seal that is used for monitoring an inventory of physical containers \cite{Humble2012}. The seal operates by transmitting a pair of photons entangled in frequency that can be generated using SPDC pumped by a narrowband laser, cf. Fig.~\ref{fig:spdc}. One of the photons is transmitted through an active fiber potentially exposed to tampering while the other passes through a reference link. For the case of nominal link behavior, both photons reach the interferometric receivers. The Mach-Zhender interferometers in these receivers operate much like the one discussed previously in Fig.~\ref{fig:MZI}. Each photon may either transmit or reflect from the first beam splitter. When the photon transmits it takes a shorter path to the next beam splitter than those that reflect and take a longer path. The two paths converge on a second beam splitter that combines the respective short-path and long-path amplitudes and forms a final superposition state. The photon is then measured with respect to its time of arrival at the detector.
\par
Quantum mechanically, the intrinsic uncertainty in the photon frequencies represents the information being transmitted. Measurements at the two receivers appear as an unpredictable series of arrival times. However, correlations between the measurement times observed at the different receivers arise because of the entanglement in the original photons. Moreover, when the photons sample the same path through the interferometer, then the joint measurements cannot distinguish between the case that both photons took the short path and the case that both photons took the long path. The indistinguishably leads to constructive interference. For the case that the photons take different paths through the interferometers, e.g., short-long or long-short, then the measurements can be used to identify exactly which case occurred and interference does not occur. 
\par
The correlations that arise in the joint measurements can be controlled through the described interference effect. A phase modulator is placed in the long path of each interferometer. If a photon travels the long path it picks up the applied phase, which can be randomly selected from a predetermined set of values. As a result, the interference between the long and short path events is modulated by the applied phase. Notably, the modulation depends on the phases applied in both interferometers and only appears when analyzing the joint measurements. The strength of the measured correlations can be quantified by the depth, or \textit{visibility}, of this interference modulation. For a maximally entangled photon pair, the modulation can have a maximal visibility of 1, while for locally correlated states the visibility has a maximum of {~0.71}. Between this threshold for non-locality and the maximum visibility is a large range of acceptable values that permits noisy components and noisy fibers to be used in constructing a quantum seal.
\par
The visibility serves as a statistic for monitoring the integrity of the first photon. If the visibility remains above the predefined threshold, then it signifies normal behavior and the authenticity of the transmission can be confirmed. However, in the presence of an intercept-resend attack on the active fiber, the entanglement between the two photons would be destroyed due to the no-cloning theorem. As a result, the visibility of a compromised fiber vanishes and the attack is readily identified. It is the restrictions based on quantum physics that provides security of the physical signal and not the speed or complexity of the transmission. Consequently, a quantum seal does not store secret information, such as a key, but is a transparent method for authenticating communication channels. Although our description of a quantum seal has focused on a particular encoding, namely, frequency entangled photons generated by SPDC, alternative realizations are also possible including weak-coherent pulse and single-photon implementations.
\section{Cyber-Physical System Security}
\label{sec:cps}
Conventional communication is a complex hierarchy of layers and protocols for synchronizing and sharing information. Securing each layer is an equally complex task with its own hierarchy of approaches. Among these, the physical layer stands out as being immune to the information theoretic approaches that characterize, for example, cryptographic security at the application or session layers. This is because the type of assurances that must be provided are very different between the physical and non-physical layers. Whereas the non-physical layers  attempt to prevent knowledge about the transmitted data, securing the physical layer requires preventing access to the transmitted signals.
\par
Preventing access to the fiber that conducts the communication signal is perhaps the most straightforward approach to securing the physical layer. However, as outlined in the Introduction, this is not a practical approach given the large footprints and impracticality of physical inspections. Complementary to restricting physical access is the use of active diagnostics, like optical time domain reflectometery (OTDR), which characterize fiber health by probing with known signal. The reflected probe pulses are measured and compared to known reference traces as a means of monitoring changes in the fiber health. The OTDR trace is exceptionally useful for diagnosing faults in fiber networks and troubleshooting connectivity problems. However, by using a classical probe signal, OTDR is vulnerable to the same man-in-the-middle attack that motivates the need for security. 
\par
The phenomenology of the quantum seal suggests a useful alternative for monitoring the integrity of a communication channel. Because the quantum seal is not vulnerable to the man-in-the-middle attack, it is possible to authenticate the transmission between two users. The authenticity of the quantum signal transmitted through the channel can be verified under the assumption of the no-cloning theorem.  By evaluating the fidelity of a received transmission against the threshold expected by Bell's test, a statistical measure of authenticity can be generated. The ability for a quantum seal to authenticate a communication link provides a new means of securing the physical layer. The diagnostic measures from a quantum seal can be part of a larger communication protocol that uses these physical measurements to determine follow-on behavior. Protocols that incorporate this type of measurement feedback into the system behavior are called \textit{cyber-physical systems}. 
\par
In the broadest sense, a cyber-physical system integrates together the computational and physical elements of a system. For our purposes, the integration of a quantum seal with a communication receiver can be viewed as a small-scale cyber-physical system. The quantum seal acts as the physical element while the associated signal processing feeds into the accompanying computational element. The diagnostic measures provided by the seal drive the behavior of this cyber-physical behavior, which can be tailored to the seal status. In the simplest setting, the cyber-physical receiver may be used to certify a burglar alarm for a physical premises or to monitor dedicated fiber-to-the-home links for unauthorized access or loss in services. On even larger scales, the seal health status of a cyber-physical receiver can be used to make  routing decisions through optical networks or for the selection of cryptographic requirements when link security is suspicious. 
\begin{figure}[h]
\begin{center}
\includegraphics[scale=0.6]{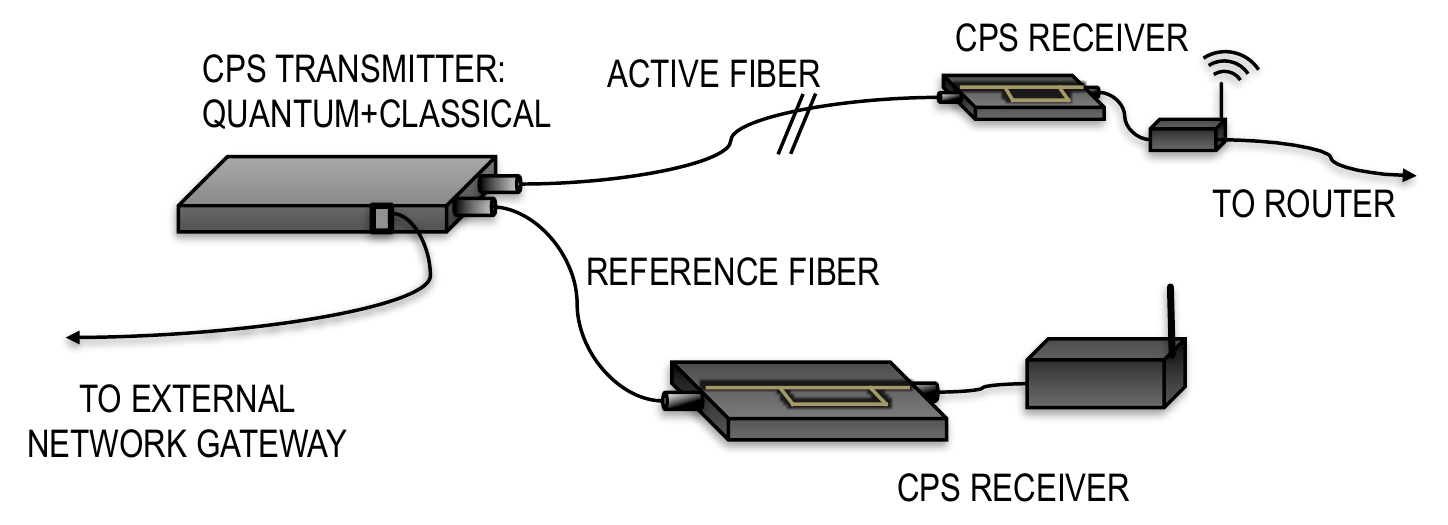}
\caption{The integration of a quantum seal with a cyber-physical monitoring system, in which the transmitter multiplexes quantum and classical signals and the receivers exchange information about the state of the seal using wireless communications. In this example, the transmitter routes packets from the network gateway through the active fiber only when the state of the seal is verified as normal. The receivers in this example may communicate observations via wireless or wired communication. Setting up the sharing of information between these different components will require standardization of the quantum physical sublayer and how those measurements are integrated into a communication system.}
\vspace{-10pt}
\label{fig:cps}
\end{center}
\end{figure}
\par
An implementation of a cyber-physical system using a quantum seal for monitoring the physical layer is shown in Fig. \ref{fig:cps}. The setup consists of a combined quantum-classical transmitter that accesses a pair of channels denoted as the active and reference links. The transmitter contains a quantum light source, like the SPDC source highlighted in Fig.~\ref{fig:seal}, to transmit one photon down the active fiber and the other down the reference link. At the end of each fiber, an interferometric receiver measures the photon arrival time just as the unbalanced Mach-Zehnder interferometer in Fig.~\ref{fig:MZI}. Local measurements are then shared with neighboring receivers in order to establish the seal health using a Bell test for non-locality.
\par
The implication of the design in Fig.~\ref{fig:cps} is that the quantum transmitter and receivers implement a quantum physical sublayer. The role of this new sublayer is to certify the integrity and authenticity of the active link. However, the physical sublayer only provides measurement feedback while seal certification requires interaction with the 'cyber' elements of the system. In this example, the transmitters and receivers must communicate local measurements to a common location, such as the original transmitter, that calculates the correlation visibility. This diagnostic is then evaluated with respect to the quantum-classical threshold to make a statement about link health. Nominal behavior is indicated when the observed correlations are above the threshold, while below threshold measurements indicates abnormal behavior and a likely attack. 
\par
When a quantum physical sublayer indicates abnormal behavior within the channel, the receiver can flag the link as being abnormal or insecure. This information can be used either for independent monitoring of fiber health or for integrating into higher protocol layers. For example, the identification of insecure links using the state of the quantum seal can serve to update policies in the data link and network layers. These higher layers can then inform decisions to the application layer, for example, about the cryptographic security necessary to establish a secure length. 
\par
The development of a quantum physical sublayer provides a useful capability for certifying the health of a fiber link. Collections of these cyber-physical system can be linked to together to form certifiably secure paths across a network. These types of networks may include communication networks, intrusion detection systems, and other sensor systems in need of authenticated signals. 
\section{Summary Outlook}
\label{sec:out}
There are several outstanding research and development issues that must be overcome before adoption of quantum physical layer security. This includes standardization of how quantum information fits into a protocol suite and what is required by a quantum physical sublayer implementation. The emergence of cyber-physical systems may offer a natural opportunity for resolving these integration issues and provide a convenient path to introducing new concepts for the physical layer. 
\par
The hardware required by a quantum physical sublayer is very similar to that called for by existing and forthcoming QKD communication systems. Both quantum seals and QKD require transmitters and receivers for single-photon or weak pulse signals. Several commercial QKD vendors, such as Id Quantique and QuintessenceLabs, already offer integrated systems containing these types of components. The main challenge for both quantum seals and cryptographic applications is the extension of these system to existing fiber networks. Recent research has shown it is possible for both quantum and classical channels to coexist \cite{Patel2012}, while the construction of larger communication architectures has been the subject of several test bed studies \cite{Peev2009,Stucki2011,Kitayama2011}.
\par
There are notable differences between monitoring the physical layer and the cryptographic security offered by QKD. Whereas QKD provides support to multiple protocol layers as part of a communication service, it is motivated by the potential for more conventional cryptography to be compromised, e..g, with the advent of quantum computers. Quantum seals, by contrast, are limited to monitoring the PHY layer and addressing the man-in-the-middle vulnerability that has already materialized. In addition, the computational and communication complexity of implementing a quantum physical sublayer is also small relative to a QKD service. For example, a quantum seal does not require the costly computations needed to generate a secret key through multiple rounds of communication like with QKD. Thus, a quantum physical sublayer may be viewed as the near-term infrastructure needed to support future QKD services.
\par
In summary, we have described an application of quantum communication for securing the physical layer that is common throughout communication protocols. The quantum seal uses the unique physical phenomenology of quantum signal to identify tampering and man-in-the-middle attacks. Like quantum cryptography, the technological basis for this security requires building and fielding transmitters and receivers that operate within the quantum regime of light.

\section*{Acknowledgment}
This work was supported by the Defense Threat Reduction Agency. This manuscript has been authored by UT-Battelle, LLC, under Contract No. DE-AC05-00OR22725 with the U.S. Department of Energy. Accordingly, the U.S. Government retains a non-exclusive, royalty-free license to publish or reproduce the published form of this contribution, or allow others to do so, for U.S. Government purposes.


\begin{thebibliography}{10}
\providecommand{\url}[1]{#1}
\csname url@samestyle\endcsname
\providecommand{\newblock}{\relax}
\providecommand{\bibinfo}[2]{#2}
\providecommand{\BIBentrySTDinterwordspacing}{\spaceskip=0pt\relax}
\providecommand{\BIBentryALTinterwordstretchfactor}{4}
\providecommand{\BIBentryALTinterwordspacing}{\spaceskip=\fontdimen2\font plus
\BIBentryALTinterwordstretchfactor\fontdimen3\font minus
  \fontdimen4\font\relax}
\providecommand{\BIBforeignlanguage}[2]{{%
\expandafter\ifx\csname l@#1\endcsname\relax
\typeout{** WARNING: IEEEtran.bst: No hyphenation pattern has been}%
\typeout{** loaded for the language `#1'. Using the pattern for}%
\typeout{** the default language instead.}%
\else
\language=\csname l@#1\endcsname
\fi
#2}}
\providecommand{\BIBdecl}{\relax}
\BIBdecl

\bibitem{Wilde2013}
M.~W.~Wilde, \emph{From Classical to Quantum Shannon Theory}.\hskip 1em plus
  0.5em minus 0.4em\relax Cambridge University Press, 2013.

\bibitem{Eisaman2011}
M.~D. Eisaman, J.~Fan, A.~Migdall, and S.~V. Polyakov, ``Invited review
  article: Single-photon sources and detectors,'' \emph{Review of Scientific
  Instruments}, vol.~82, no.~7, p. 071101, 2011.

\bibitem{Johnston2001}
R.~G. Johnston, ``Tamper detection for safeguards and treaty monitoring:
  Fantasies, realities, and potentials,'' \emph{The Nonproliferation Review},
  vol. 8.1, p. 102, Spring 2001.

\bibitem{Szustakowski2005}
M.~Szustakowski and M.~Zyczkowski, ``Fiber optic sensors for perimeter security
  with intruder localisation,'' \emph{Proc. SPIE}, vol. 5954, p. 59540C, 2005.

\bibitem{Juarez2005}
J.~Juarez, E.~Maier, K.~N. Choi, and H.~Taylor, ``Distributed fiber-optic
  intrusion sensor system,'' \emph{Journal of Lightwave Technology}, vol.~23,
  no.~6, pp. 2081 -- 2087, June 2005.

\bibitem{Wootters1982}
W.~K. Wootters and W.~H. Zurek, ``A single quantum cannot be cloned,''
  \emph{Nature}, vol. 299, pp. 802--803, 1982.

\bibitem{Humble2012}
T.~Humble, D.~Earl, and B.~Williams, ``Tamper-indicating quantum optical
  seals,'' in \emph{Photonics Conference (IPC)}, pp. 475--476. (2012)

\bibitem{Patel2012}
K.~A. Patel et~al., ``Coexistence of high-bit-rate quantum key
  distribution and data on optical fiber,'' \emph{Phys. Rev. X}, vol.~2, p.
  041010, Nov 2012.

\bibitem{Peev2009}
M. Peev et~al., ``The {SECOQC} quantum key distribution network in {V}ienna,''
  \emph{New Journal of Physics}, vol.~11, no.~7, p. 075001, 2009.

\bibitem{Stucki2011}
D. Stucki et~al., ``Long-term performance of the {S}wiss{Q}uantum quantum key
  distribution network in a field environment,'' \emph{New Journal of Physics},
  vol.~13, no.~12, p. 123001, 2011.

\bibitem{Kitayama2011}
K.-I. Kitayama et al., ``Security in photonic networks:
  Threats and security enhancement,'' \emph{Journal of Lightwave Technology},
  vol.~29, no.~21, pp. 3210--3222, November 2011.

\end{thebibliography}

\end{document}